\documentclass[review]{elsarticle}

\usepackage{lineno,hyperref}
\modulolinenumbers[5]
\usepackage{color}
\usepackage{ulem}
\usepackage{tabularx,ragged2e,booktabs,caption,graphicx,epstopdf,float}

\journal{Carbon}
\begin{document}
\begin{frontmatter}

\title{Zeta potential and nanodiamond self assembly assisted diamond growth on lithium niobate and lithium tantalate single crystal}

\author[1]{Soumen Mandal\corref{a}}
\ead{mandals2@cardiff.ac.uk}
\author[2]{Karsten Arts}
\author[3]{David Morgan}
\author[4]{Zhuohui Chen}
\author[1]{Oliver A. Williams\corref{a}}
\ead{williamso@cardiff.ac.uk}
\address[1]{School of Physics and Astronomy, Cardiff University, Cardiff, UK}
\address[2]{Eindhoven University of Technology, 5612 AZ, Eindhoven, the Netherlands}
\address[3]{Cardiff Catalysis Institute, School of Chemistry, Cardiff University, Cardiff CF10 3AT, U.K.}
\address[4]{Huawei Technologies Canada Co., Ltd., 303 Terry Fox Drive, Kanata, Ontario, K2K 3J1, Canada}
\cortext[a]{Corresponding authors}

\begin{abstract}
This study focuses on the self-assembly and subsequent diamond growth on SiO$_2$ buffered lithium niobate (LiNbO$_3$) and lithium tantalate (LiTaO$_3$) single crystals. The zeta-potential of LNO and LTO single crystal were measured as a function of pH. They were found to be negative in the pH range 3.5-9.5. The isoelectric point for LNO was found to be at pH $\sim$ 2.91 and that of LTO to be at pH $\sim$ 3.20. X-ray photoelectron spectroscopy performed on the surfaces show presence of oxygen groups which may be responsible for the negative zeta potential of the crystals. Self-assembly of nanodiamond particles on LTO and LNO, using nanodiamond colloid, were studied. As expected, high nanodiamond density was seen when self-assembly was done using a positively charged nanodiamond particles. Diamond growth was attempted on the nanodiamond coated substrates but they were found to be unsuitable for direct growth due to disintegration of substrates in diamond growth conditions.. A $\sim$100nm thick silicon dioxide layer was deposited on the crystals, followed by nanodiamond self assembly and diamond growth. Thin diamond films were successfully grown on both coated crystals. The diamond quality was analysed by Raman spectroscopy and atomic force microscopy.
\end{abstract}


\end{frontmatter}


\section{Introduction}
Sound waves propagate in air through mechanical vibrations which happens along the direction of propagation. Similar effects can also be seen in solids when a wave passes through the solid. The phonons which represent movement of atoms in the lattice along the direction of propagation of wave are also known as acoustic phonons (like sound wave propagating in a media). Like phonons, another particle of interest to many applications are the photons. They have striking similarity in the form that sound in rigid media and light in transparent objects can travel without attenuation. However, the higher speed of light ($\sim$ 100000 times that of sound) means that photonics devices have large physical dimensions due to large wavelengths (centimetre to meters depending on frequency). Replacing light by sound makes it possible to reduce the size of devices. One such device of interest based on acoustic phonons is the surface acoustic wave (SAW) device which play a major role in variety of industrial and academic applications\cite{dels19}. For example, SAW filters are essential components for information and telecommunication systems\cite{ieee, bal22}. Conventionally SAW filters are fabricated on piezoelectric materials like lithium tantalate (LTO), lithium niobate (LNO) etc. The 5G bands proposed in the 5G road map\cite{ieee} incorporates frequency bands above 10 GHz. This puts higher power and frequency handling demand on SAW filters\cite{bal22, shen22}.  The higher operation frequency can be achieved by increasing acoustic wave velocity, which is material dependent, or by reducing the wavelength by reducing the pitch of the electrodes\cite{bal22}.  The first method requires research on newer materials with high acoustic wave velocity, the second method is limited by the lithography techniques and handling of the heat generated by the electrodes fabricated on the current piezo materials like LTO and LNO. The power handling capacity of these materials can be enhanced by attaching these materials to high thermal conductivity material like diamond ($\sim$ 1200 - 2000 Wm$^{-1}$K$^{-1}$)\cite{iva02}. This can be achieved in two ways. The first is to put the diamond on the piezo material either by direct growth\cite{hee13, man19} or by direct bonding\cite{tak99, tor08, mu18, lian19, mats20} and the second, is grow the piezo electric material on top of polished diamond\cite{rod12, mig22}. The growth of piezo material on diamond leads formation of polycrystalline piezo material which has been shown to effect device performance when compared with single crystal material\cite{mig22}. Furthermore, the current state-of-the-art for direct bonding is limited to single crystal diamond. The small sizes and high cost of single crystal diamond makes this technology unattractive. That leaves with one alternative, that is to grow a thick diamond layer on top of the piezo material. It has been shown earlier that to take full advantage of superior thermal conductivity of diamond, the diamond layers should be $\geq 10 \mu$m in thickness\cite{zha21}.

Diamond growth on non-diamond substrates is non-trivial process due to large differences in surface energies between diamond and most substrates. With a surface energy of $\sim$ 6J/m$^2$\cite{hark42}, diamond has much higher surface energy than common substrates like silicon ($\sim$ 1.5J/m$^2$)\cite{jacc63}. This energy difference leads to isolated island like growth of diamond crystals on substrates\cite{will11rev, mandrev} like silicon with an island density of around 10$^4$ - 10$^5$ cm$^{-2}$. The surface energy for LNO\cite{hir20} is $\sim$ 1.1J/m$^2$. These energies were calculated using the cleavage technique\cite{obr30}. For LTO, such cleavage based surface energies are not present, however, surface energies calculated based on water contact angle shows LTO to have similar surface energies to silicon\cite{bak19}. Considering the large differences in surface energies, for growth of diamond on these substrates they need to be coated with small diamond particles. For a smooth interface the particles need to be as small as possible. The smallest such particles are admantane and the process of coating such small particles on any surface is quite involved\cite{man16}. The easier approach would be to self assemble slightly bigger particles on to the substrate surfaces using nanodiamond ($\sim$ 5-10 nm in size) colloid\cite{will07s, hee11}. However, the self-assembly of nanodiamonds is only the first step in the diamond growth process. The coated substrates are exposed to harsh diamond growth conditions in the presence of hydrogen plasma. Normally, the growth temperatures for diamond is of the order of 800 $^o$C\cite{will11rev}, which is higher than the Curie temperature of LTO\cite{lev66}. It has also been found in the case of LNO that the crystal structure of the surface is completely destroyed/blackened when exposed to hydrogen plasma\cite{tur98}. This can be recovered to some extent by exposing to oxygen plasma, however, if diamond is grown on the surface then the LNO substrate is completely blanketed and no diffusion of oxygen can happen to reform the surface. Similar, blackening effect was also observed for LTO substrates, during this study, when exposed to diamond growth conditions. Furthermore, the large difference in  thermal expansion coefficient between the substrates\cite{kim69} and diamond\cite{moel97} means that the substrate temperature has to be kept as low as possible to manage stress in the film when cooled to room temperature after growth. Considering all the above, it was decided to have similar approach to $\beta$-Ga$_2$O$_3$ for the growth of diamond on LNO and LTO\cite{man21}. The substrates were coated with 100nm of SiO$_2$ by atomic layer deposition (ALD) before self assembly of nanodiamond and growth.  Diamond growth on LNO has been attempted before\cite{jag04}, however the seeding technique used for the growth was mechanical scratching. It is well known that mechanical scratching can lead to rough interface between diamond and substrate\cite{will11rev, mandrev}. Such rough interfaces have been shown to adversely affect SAW performances in the past\cite{vor11}.

To summarise, in this work, 1) the zeta($\zeta$)-potentials of LNO and LTO were measured, 2) self assembly of nanodiamond was studied on LNO and LTO using AFM, 3) the substrates were coated with silicon dioxide and diamond growth was attempted, and 4) the diamond layer on the substrates were examined with Raman spectroscopy and atomic force microscopy.

\section{Methods and Materials}
The $\zeta$-potential of LNO and LTO surfaces were measured using Anton Paar Surpass$^{TM}$ 3. The instrument measures the $\zeta$-potential by measuring the streaming potential, while passing an electrolyte, between the inlet and outlet of a narrow channel formed by the sample surfaces\cite{wag80}. The streaming potential is a result of moving counterions which are sheared off the sample surfaces by the passing electrolyte\cite{man21}. The measurement of streaming potential as a function of electrolyte pressure can give the $\zeta$-potential through Helmholtz-Smoluchowski equation\cite{wern98}. The electrolyte used in this study was 10$^{-3}$ M KCl solution, the channel width was kept between 90 and 110 $\mu$m and the electrolyte pressure was varied between 200 and 600 mbar. The surface charge of any solid in a liquid is heavily dependent on the pH of the solution which in turn effects the $\zeta$-potential. To vary the pH during the measurement, 0.05M NaOH and 0.1M HCl solution was used with the inbuilt titrator in Surpass$^{TM}$ 3. Hydrochloric acid was sourced from Fisher chemicals (Product code: H/1150/PB15), sodium hydroxide (Product code: 28245.265) and potassium chloride (Product code: 26764.260) were purchased from VWR. The DI water was from ReAgent Chemical Services.

The knowledge of $\zeta$-potential helps in choosing the right kind of nanodiamond colloid (positive or negative) for electrostatically driven self assembly of nanodiamonds on LNO and LTO surfaces\cite{will07s, will10h, will11rev, mandrev}. Two different types of suspensions were used in this work. The positively charged suspension was made using hydrogen terminated nanodiamonds with average particle size of 10nm and $\zeta$-potential of +40mV. The negatively charged suspension was prepared with oxygen terminated nanodiamonds and had an average particle size of 10nm with -50mV $\zeta$-potential. Full details of making the suspensions have been published elsewhere\cite{will10h, will11rev, hee11, mandrev, blan21}. While self-assembly of nanodiamond on LNO and LTO is straightforward with knowledge of surface charge, the growth of diamond thin film from the self-assembled particles is non-trivial. The growth of diamond on these substrates, like Ga$_2$O$_3$\cite{man21}, is dependent on the ability of the substrates to withstand extreme diamond growth conditions as well as the difference in coefficient of thermal expansion between the substrates and diamond. This is due to the normal growth conditions of diamond being 500$^o$ C or higher\cite{will11rev}. To work around this problem a $\sim$100nm thick silicon dioxide layer was grown by atomic layer deposition (ALD). An Oxford Instruments FlexAL reactor was used for this process\cite{heil07} with 1100 ALD cycles with a table temperature of 300 $^o$C. SiH$_2$(NEt$_2$)$_2$ (bis(diethylamino)silane) precursor was used along with O$_2$ plasma, generated by an inductively coupled plasma source, as coreactant. The thickness of the SiO$_2$ layer was measured by J.A. Woollam M-2000D spectral ellipsometer\cite{lang09}. For the thickness measurement a small piece of silicon substrate was introduced at the same time as the LTO and LNO substrates. Previous measurements of $\zeta$-potential on ALD grown SiO$_2$ have shown the surfaces to be negatively charged in water (have negative zeta potential)\cite{man21}.

For studying the self assembly of nanodiamond on LNO and LTO surfaces, atomic force microscopy (AFM) images were taken using Bruker Dimension Icon atomic force microscope. The microscope was operating in peak force tapping mode using a ScanAsyst tip. For this study three pieces of each substrate were taken. One piece from each type of substrate was dipped in diamond colloid containing positive nanodiamond particles. The substrates were sonicated in the colloid for 10 mins and then spin dried at 3000 rpm. The same was repeated with a second set of pieces by dipping them in diamond colloid containing negatively charged particles. The oxide coated substrates were dipped in positively charged nanodiamond colloid for self-assembly of diamond particles. The particle loaded substrates were then introduced in a clamshell microwave chemical vapour deposition system(CVD)\cite{man17}. Growth on both substrates were done at 5\% CH$_4$/H$_2$ ratio, with CH$_4$ acting as carbon source and H$_2$ acting as etchant for non-diamond carbon\cite{will11rev, cuen22}. However, the growth conditions were slightly different due to difference in substrate thickness. The main aim was to keep the substrate temperature close to 500 $^o$C to avoid excessive stress between grown diamond and substrates. The stress is due to difference in coefficient of thermal expansion between diamond\cite{moel97} and the substrate\cite{kim69, cuen21}. For the thinner coated-LTO substrate the growth conditions were 3.2kW of microwave power with 35 torr gas pressure. The coated-LNO substrates were subjected to 3.2 kW of microwave power under 29 torr gas pressure. The lower pressure reduces slightly the overall energy density of the plasma, thus giving lower substrate temperature for the thicker substrate\cite{cuen22}. Since the difference between the growth conditions is small, the overall growth chemistry is not significantly affected to result in diamond layers with dissimilar properties\cite{cuen22}. The diamond films on both substrates were grown for 120 mins and cooled to room temperature in hydrogen plasma. Following the growth of diamond film was analysed using a SynapsePlus Back-Illuminated Deep Depletion (BIDD) CCD equipped Horiba LabRAM HR Evolution. The excitation wavelength was 532 nm.  The surfaces of the films were imaged with atomic force microscope. X-cut LTO and LNO were commercially sourced. The LTO substrates were 5 X 5 mm in size and 0.5 mm in thickness. The LNO substrates were 5 X 5 mm in size and 1 mm in thickness. The substrate surface were examined with X-ray photoelectron spectroscopy (XPS). The data was collected using a Thermo Scientific K-Alpha$^+$ spectrometer with a monochromatic Al source that was operated at 72W, with an emission current of 6mA at a 12kV anode potential. Survey and high-resolution spectra were obtained using pass energies of 150 and 40 eV, respectively. To neutralize the charge, a combination of electrons and low energy Ar ions were employed. The spectra were analyzed using CasaXPS 2.3.26\cite{fair21} software.

\section{Results and Discussion}
\begin{figure}[t]
\centering
\includegraphics[width = 9cm]{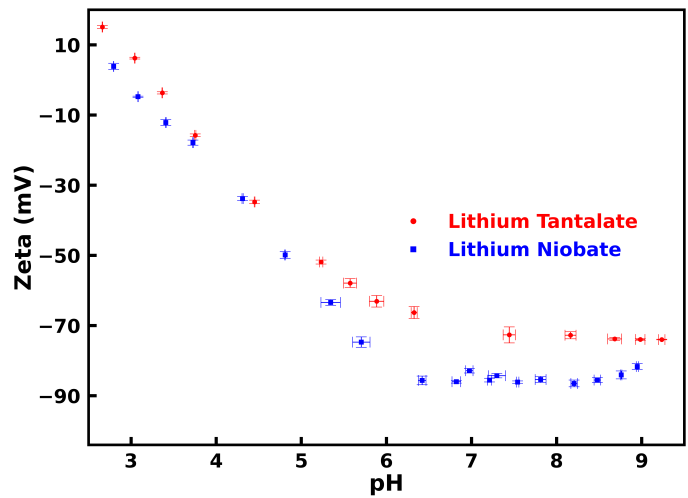}
\caption{$\zeta$-potential of lithium tantalate and lithium niobate single crystal as a function of electrolyte pH. The $\zeta$-potential for both the crystals are negative in the pH range 3.5-9.5 with isoelectric points close to pH $\sim$3.} \label{fig1}
\end{figure}

Figure \ref{fig1} shows the $\zeta$-potential of LNO and LTO single crystal as a function of pH. The $\zeta$-potential is negative in the pH range 3.5-9.5 for both the substrates. The isoelectric point for LNO is around the pH value of $\sim$2.91 while that for LTO is $\sim$3.20. The isoelectric points were calculated by fitting a trend-line to the curves around 0 mV mark. This information is critical for determining the type of nanodiamond suspension needed for high density nanaodiamond self assembly on the substrates\cite{will11rev, mandrev, hee11}. It has been shown that particle in nanodiamond colloids can be positively or negatively charged depending on the pre-treatment of the particles used for colloid formation\cite{hee11}. Since the self assembly is electrostatically driven, nanodiamond particles with opposite $\zeta$-potential to substrate are best suited for high density self assembly\cite{gira09, scor09}. In general the pH of the nanodiamond colloid is around 4-6 and in this region the $\zeta$-potential of LNO and LTO substrates are negative (Fig \ref{fig1}). So, for a high-density of nanodiamond particle on the substrate surface a positively charged colloid is best suited. 

\begin{figure}[t]
\centering
\includegraphics[width = 9cm]{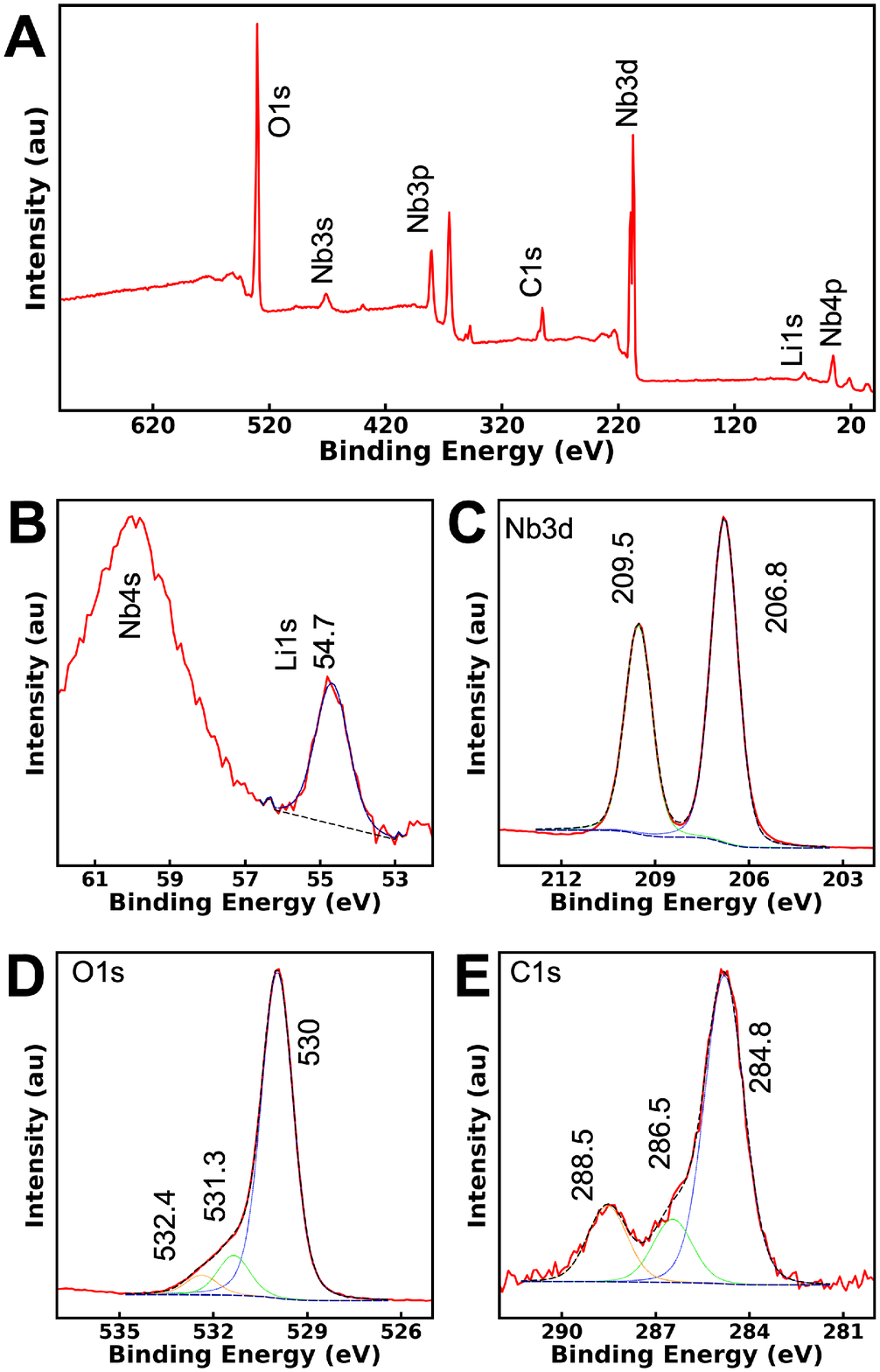}
\caption{XPS spectra of LNO surface. Panel A shows the survey spectra of the sample. The standard peaks for the crystal are indicated next to the peaks. Panel B shows the zoomed in view of the Li1s peak region. Panel C is the zoom in of the Nb3d peaks. Panels D and E shows the O1s and C1s peaks respectively.} \label{xpsnb} 
\end{figure}

\begin{figure}[t]
\centering
\includegraphics[width = 9cm]{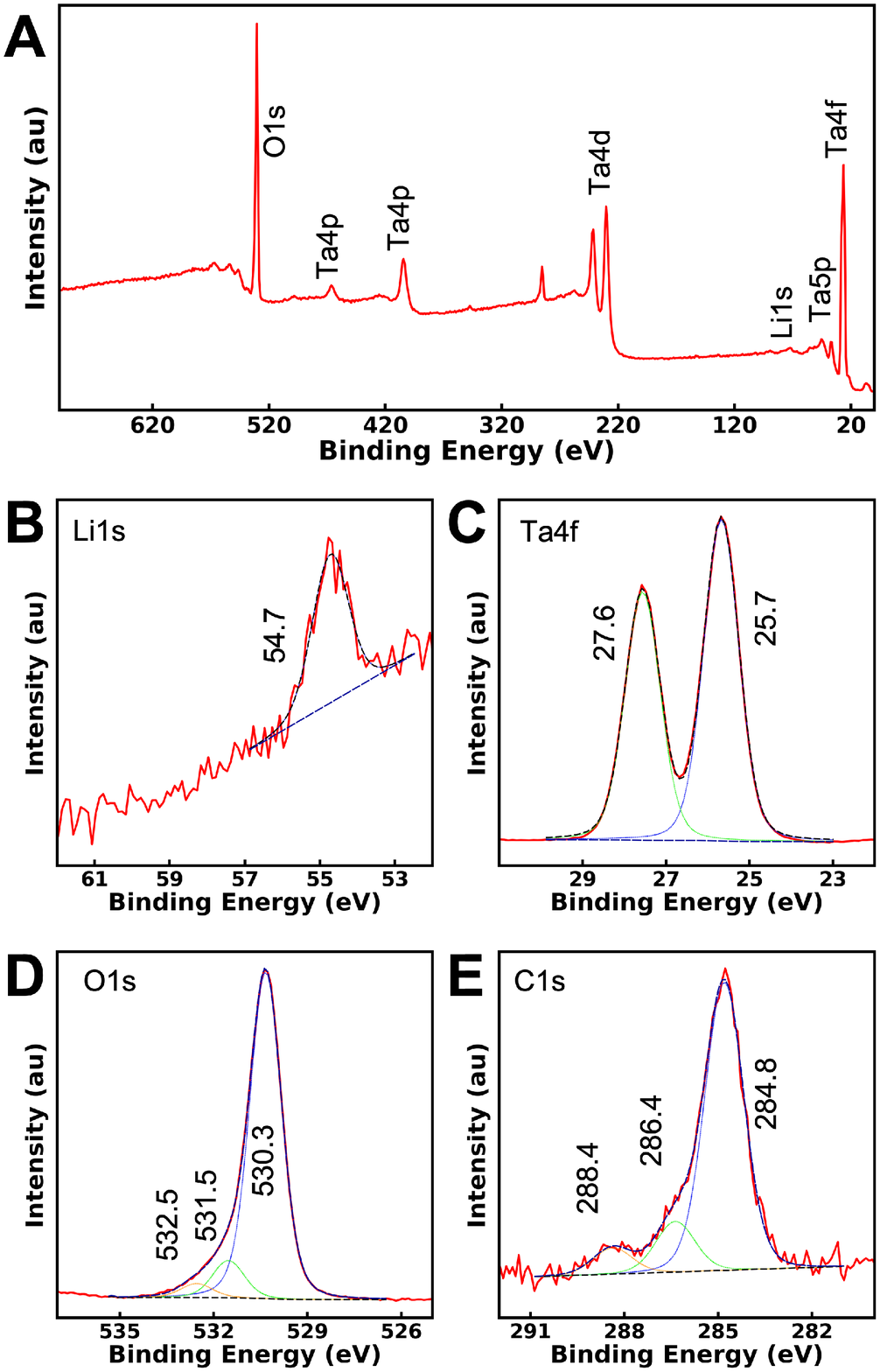}
\caption{The XPS survey spectra of LTO surface is shown in Panel A, with the standard peaks for the crystal indicated next to the peaks. Panel B displays a closer view of the Li1s peak region, while Panel C is a magnified view of the Ta4f peaks. Panels D and E present the O1s and C1s peaks, respectively.} \label{xpsta}
\end{figure}

A negative zeta potential in oxides is typically observed due to the adsorption of negatively charged species on the oxide surface, such as hydroxyl groups (OH-) or other anions\cite{hunt81}. XPS was done on both LNO and LTO surfaces to examine the crystal surfaces. The spectra for LNO and LTO are presented in figures \ref{xpsnb} and \ref{xpsta} respectively. The charge correction for the spectra were done using the 284.8eV adventitious carbon peak\cite{wilso01, ferr05}. The standard peaks in both LNO\cite{skry16} and LTO\cite{gitm95, ono02} crystals are indicated in panel A of figures \ref{xpsnb} and \ref{xpsta} respectively. Figure \ref{xpsnb}B shows the zoomed in region of the Li1s peak. The peak position as determined by fitting the data using CasaXPS was found to be 54.7 $\pm$ 0.1eV. The Nb4s peak at $\sim$60 eV is also visible. Panel C shows the  Nb3d peaks corresponding to 3d$_{5/2}$ and 3d$_{3/2}$\cite{yun07}. The 3d$_{5/2}$ peak is at 206.8 $\pm$ 0.1 eV and the 3d$_{3/2}$ peak is at 209.5 $\pm$ 0.1 eV. The difference in the peak position is around $\sim$ 3eV which is consistent with what has been seen in the literature\cite{yun07} for z-cut LNO. The region of the spectra of particular interest for purpose of negative $\zeta$-potential are the O1s and C1s peaks. These are shown in figure \ref{xpsnb}D and E respectively. Looking at the O1s peaks three clear peaks can be seen at 530 $\pm$ 0.1 eV, 531.3 $\pm$ 0.1 eV and 532.4 $\pm$ 0.1eV. The peak at 530 $\pm$ 0.1 eV corresponds to structural oxygen and the peak at 531.3 $\pm$ 0.1 and 532.4 $\pm$ 0.1 eV corresponds to adsorbed oxygen\cite{skry16}. The bonding state of the adsorbed oxygen atoms can be estimated by looking at the C1s peak from the crystal. Since carbon does not form part of the LNO crystal, all the carbon atoms detected on the surface are assumed to be adsorbed carbon atoms. The C1s peak can be deconvoluted into three peaks. The peak at 284.8 eV (also the reference peak for charge correction) is the sp$^3$ C-C bond\cite{wilso01, ferr05}. The second peak is at 286.5 $\pm$ 0.1 eV which is $\sim$1.7 eV away from C-C peak. This is attributed to C-OH bonds\cite{nots99, nots00} confirmed also by 531.3 eV peak in O1s spectra. Finally the third peak is at 288.5 $\pm$ 0.1 eV, which is $\sim$3.7 eV higher than C-C peak position at 284.8 eV. This is generally attributed to carbonyl (C=O) groups on the surface\cite{zaj01} which is also confirmed by the 532.4 eV peak in O1s spectra\cite{gard95}. The dashed lines at the bottom of each zoomed in region shows the inelastic background. Clearly, there is irrefutable evidence of oxygen groups (other than structural oxygen) on the surface. As has been stated before these groups are mostly responsible for the negative $\zeta$-potential in solid surfaces\cite{hunt81}.

Similar analysis can be done for the XPS spectra for LTO crystal. Figure \ref{xpsta}B shows the Li1s peak at 54.7 $\pm$ 0.1 eV\cite{ono02}. Figure \ref{xpsta}C shows the Ta4f peaks at 27.6 and 25.7 $\pm$ 0.1 eV which correspond to oxidised Ta4f$_{5/2}$ and Ta4f$_{7/2}$ states respectively\cite{gitm95, zhen20}. A very small hump seen close to 22 eV can signal the presence of some unoxidised Ta on the surface\cite{gitm95}. Panels D and E in figure \ref{xpsta} shows the spectra for O1s and C1s peaks, respectively, from LTO surface. Similar to LNO, the O1s peak of LTO can be convoluted into three peaks at 530.3, 531.5 and 532.5 $\pm$ 0.1 eV. The peak at 530.3 eV corresponds to structural oxygen (similar to LNO) and the peak at 531.5 and 532.5 eV corresponds to adsorbed oxygen\cite{zhen20}. Finally, looking at the C1s peak of the LTO surface, a picture similar to LNO surface appears. The C1s peak exhibits three distinct peaks upon deconvolution. The first peak at 284.8 eV is considered as the sp$^3$ C-C bond and serves as the reference peak for charge correction, as supported by previous studies\cite{wilso01, ferr05}. The second peak, located at 286.4 $\pm$ 0.1 eV, is approximately 1.8 eV away from the C-C peak and is associated with the presence of C-OH bonds\cite{nots99, nots00}. The third peak, which appears at 288.4 $\pm$ 0.1 eV, is about 3.6 eV higher than the C-C peak at 284.8 eV, and is commonly attributed to carbonyl (C=O) groups present on the surface\cite{zaj01}. This observation is further confirmed by the 531.5 and 532.5 eV peaks observed in the O1s spectra\cite{gard95}. It is evident that oxygen groups, besides structural oxygen, exist on the surface of LTO substrate similar to LNO. As has been stated earlier, these groups are primarily responsible for the negative $\zeta$-potential exhibited by solid surfaces\cite{hunt81}. The LNO crystal had trace amounts of sodium, copper, sulphur, calcium, cobalt and silicon on the surface, the origin for which is not known. Similarly, the LTO substrates had trace amounts of zinc, sodium and cobalt. There are still some subtle difference in the $\zeta$-potential of LNO and LTO surfaces which cannot be explained by the XPS study alone. Further studies are required by preferentially terminating the surfaces with specific surface groups to pinpoint the true effects of the surface group on $\zeta$-potential. Such studies are beyond the scope of the current manuscript.

\begin{figure}[t]
\centering
\includegraphics[width = 9cm]{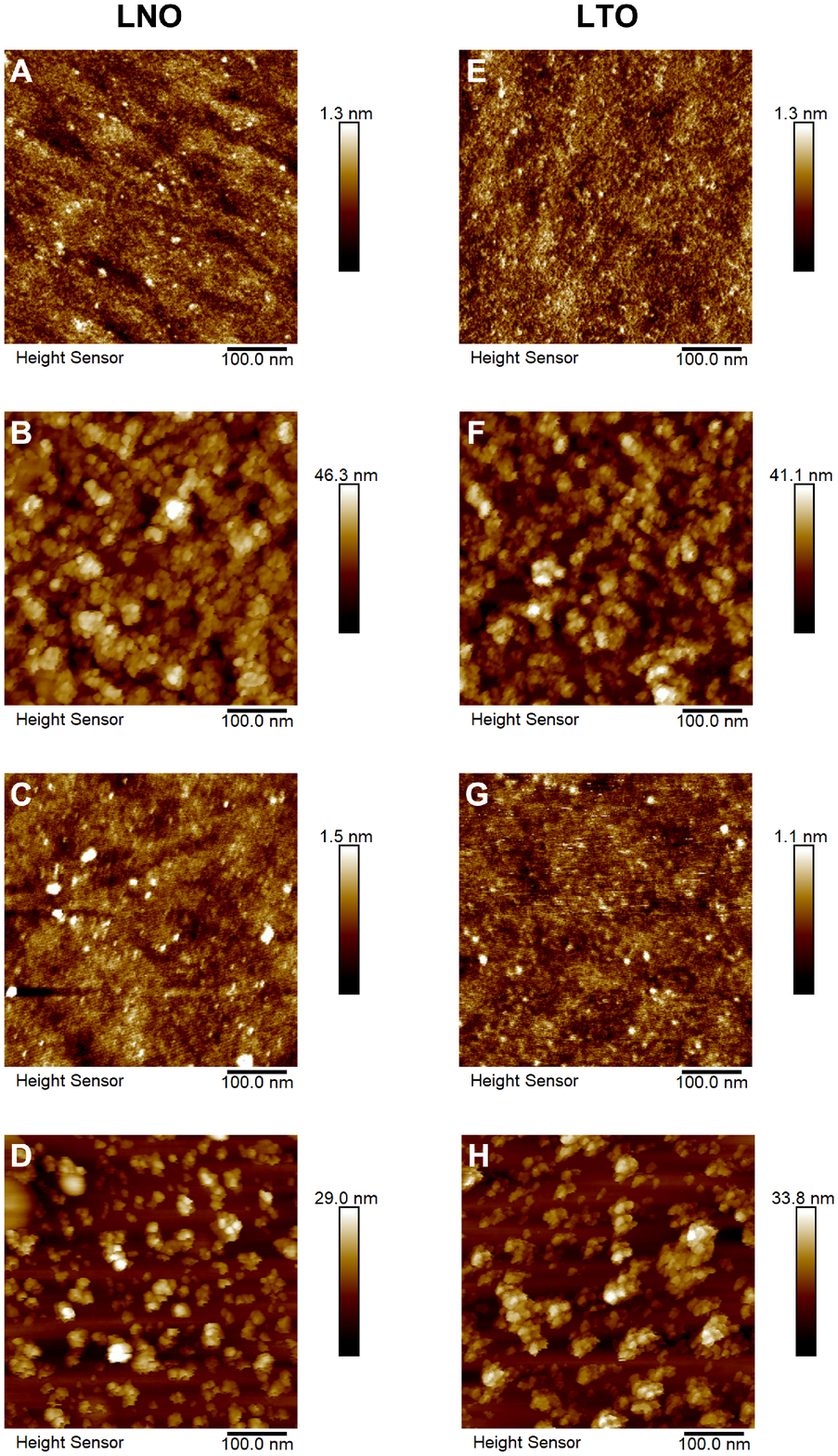}
\caption{AFM images showing self-assembly of diamond nanoparticles on LNO and LTO surfaces. The left side images  (Panel A, B, C and D) are for LNO and the right side images (Panels E, F, G and H) are for LTO. Panels A and E show the bare solvent cleaned substrate. Panels B and F for substrates dipped in positively charged diamond colloid. Finally, panels C and G show substrates dipped in negatively charged diamond colloid. Finally panels D and H are for silicon dioxide coated LNO and LTO substrates treated with positively charged diamond colloid.} \label{fig2}
\end{figure}

The substrates were dipped in nanodiamond suspensions(positive and negative) and the coated surfaces were examined using AFM. The AFM images of the surfaces (uncoated and coated) are presented in figure \ref{fig2}. Panel A, B, C, E, F, and G are for the bare substrates. Panels D and H are for SLTO and SLNO respectively. The two columns are marked on the top with the base substrate only. Panels A and E show the image of extremely smooth LNO and LTO substrate surfaces respectively. The scale bar on the side of the images gives an idea of the variation in the surface. Panels B and F in the same figure show the substrates treated with positively charged nanodiamond colloid. Evidence of self assemble of nanodiamond is clearly visible. It is evident that there are some agglomerates in the colloid which get deposited on to the substrates. In contrast when the substrates are dipped in a colloid containing negatively charged particles there is hardly any change in the roughness of the substrate surface and this is shown in figure \ref{fig2}C and G. A comparison between scale bars of panels A and C for LNO (D and F for LTO) clearly shows no self-assembly of nanoparticles on the substrates.  Finally the AFM images of the SLTO and SLNO substrates dipped in positively charged diamond colloids are shown in panels D and H. Clear self-assembly of nanodiamond are visible. However, when compared with bare LTO and LNO the seed density is slightly lower. This is due to slightly lower negative $\zeta$-potential (absolute value) of the ALD deposited silicon surface when compared with bare substrates (LTO/LNO)\cite{man21}.

The substrates with the nanodiamond particles were then introduced in a CVD reactor to directly grow diamond. As has been seen in the case of Ga$_2$O$_3$\cite{man21}, both LNO and LTO substrates were heavily etched in the hydrogen plasma thus inhibiting growth of diamond. As a result SiO$_2$ coated substrates were used to self-assemble diamond nanoparticles followed diamond growth process. It has been shown in the past that ALD grown SiO$_2$ layer have negative charge when dipped in water\cite{man21}. As a result positively charged diamond solution was used to achieve high nanodiamond density on the substrates\cite{hee11, man21}.  The diamond films were imaged using AFM and their quality tested using Raman spectroscopy. In general scanning electron microscopy(SEM) is used for imaging diamond layers, however on this occasion due to highly insulating nature of the substrate and the resulting diamond it was not possible to use SEM due to excessive charge accumulation on the samples.

\begin{figure}[t]
\centering
\includegraphics[width = 9cm]{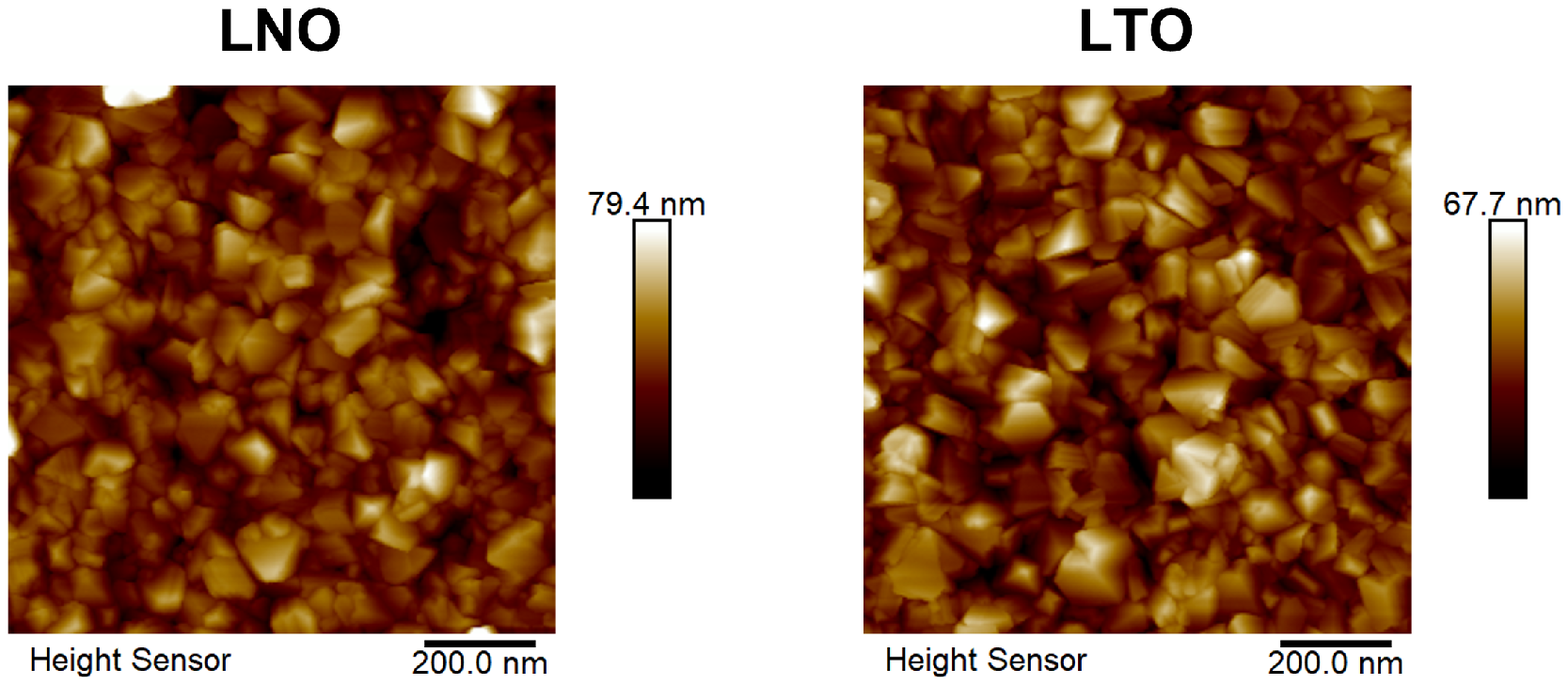}
\caption{AFM images of the thin diamond films grown on SLNO and SLTO. The crystalline facets of the small diamond grains are clearly visible with average grain size of  $\sim$ 80nm for both samples.} \label{fig3}
\end{figure}

\begin{figure}[t]
\centering
\includegraphics[height = 0.75\textheight]{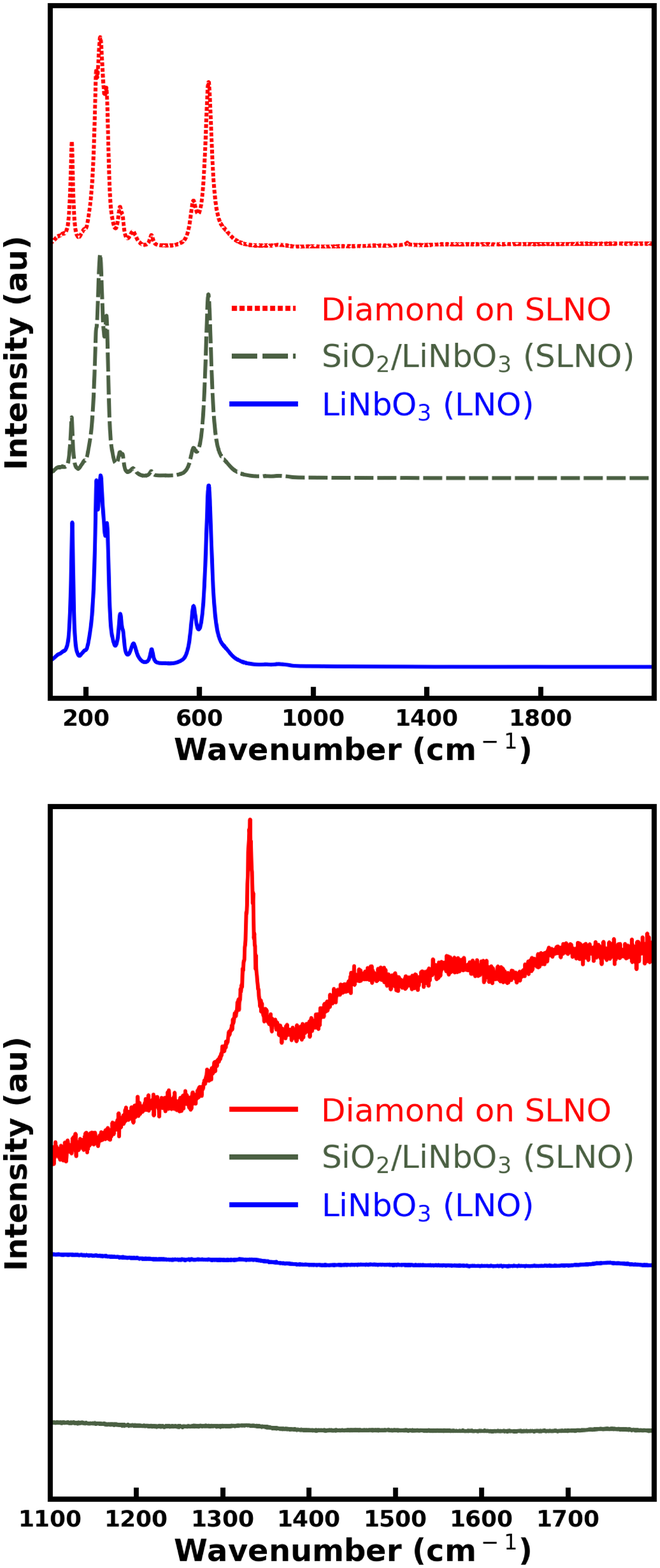}
\caption{Raman spectroscopy of the LNO substrate, SiO$_2$ coated LNO (SLNO) and the thin diamond film grown on SLNO are shown here. In all three spectra the peaks from the LNO are clearly seen. The bottom panel shows the zoomed in region between 1100 and 1800 cm$^{-1}$ clearly showing a Raman peak around 1332 cm$^{-1}$.} \label{fig4}
\end{figure}

\begin{figure}[t]
\centering
\includegraphics[width = 9cm]{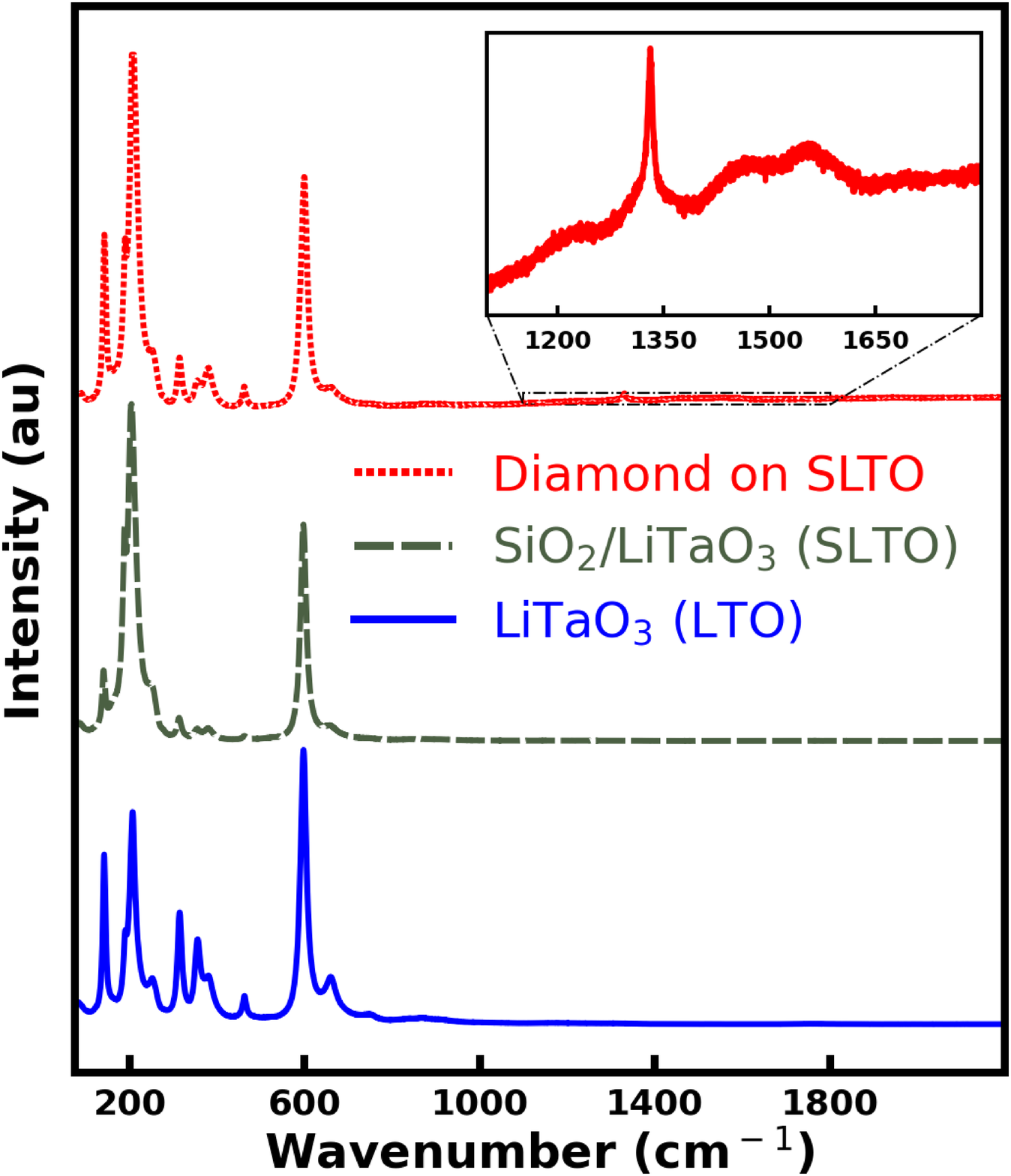}
\caption{Raman spectroscopy of the LTO substrate, SiO$_2$ coated LTO (SLTO) and the thin diamond film grown on SLTO are shown here. In all three spectra the peaks from the LTO are clearly seen. The inset shows the zoomed in region between 1100 and 1800 cm$^{-1}$ clearly showing a Raman peak around 1332 cm$^{-1}$.} \label{fig5}
\end{figure}

The AFM images of thin diamond films are shown in figure \ref{fig3}. The films grown on both the substrate show good crystalline quality as clear diamond facets are visible in the images. The presence of well faceted crystals indicate good quality of the diamond film. The average grain size is $\sim$ 80nm for both the films as seen from AFM images. The growth of diamond form nanoparticle follows a Volmer-Weber growth model\cite{vol26, jia94} until coalescence and then follows a competitive columnar growth as explained in Van der Drift model\cite{van67, smer05}. Assuming there is no secondary nucleation, the thickness of the films are close to the grain size seen at the surface. In this case, the films thickness as seen from grain size is close to $\sim$ 80nm. Considering the growth was done for 120mins, the growth rate of diamond on these substrates is of the order of $\sim$40 nm/hr. The thin films were then analysed using Raman spectroscopy and the results for the same are presented in figures \ref{fig4} and \ref{fig5}. 

Figure \ref{fig4} shows the Raman spectra of LNO, SiO$_2$ coated LNO (SLNO) and diamond film grown on SLNO. The top panel shows the full spectrum while the bottom panel shows the zoomed in view of the spectra between 1100 and 1800 m$^{-1}$. The curve shown in blue solid line is for the bare substrate. The substrate shows all the characteristic peak for LNO grown with lithium-7 ($^7$Li)\cite{sch66, rep99}. The observed peaks, as reported in literature\cite{rep99}, are 155, 180, 238, 255, 265, 276, 325, 334, 371, 431, 582, 633 and 610 cm$^{-1}$. In this case the 180 cm$^{-1}$ is not present, instead a small shoulder can be seen at $\sim 196$ cm$^{-1}$. This peak have been theoretically calculated by Repelin etal.\cite{rep99} in their work and can appear at 180 cm$^{-1}$ instead. Furthermore, the peaks at 255 and 265 cm$^{-1}$ appears as one broad band around $\sim 250$ cm$^{-1}$. The curve in green dashed line in the figure shows the spectrum for SLNO. In this case the relative intensities between various peaks are markedly different from the relative intensities of LNO spectrum. This has been seen earlier in the case of $\beta$-Ga$_2$O$_3$\cite{man21}. Finally, the data for thin diamond film grown on SLNO is shown at the top in red dotted lines. The data shows all the peaks characteristic of LNO. The signature of the thin diamond film is not visible in the zoomed out view. This is due to thin diamond film and low signal intensity when compared with the LNO single crystal substrate. The bottom panel in figure \ref{fig4} shows the expanded view around the 1332 cm$^{-1}$. A clear diamond peak can be seen in the inset along with some signature for D and G peaks. No such peaks could be seen for the Raman spectra on LNO and SLNO. The diamond peak position as determined by a Gaussian fit is $\sim$ 1331.4 cm$^{-1}$. Based on various models to calculate stress from Raman shift ($\sim$0.4GPa/cm$^{-1}$) and the ideal diamond peak at 1332 cm$^{-1}$\cite{ram30, bhag30}, the stress in the film is around $\sim$0.24 GPa\cite{bop85, knig89, anas99}. The origin of the stress in the diamond films is the difference in coefficient of expansion between substrate\cite{kim69} and diamond\cite{moel97}. Higher the growth temperature, higher would be the stress in the diamond film once the film is cooled to room temperature from growth temperature (normally around 800 $^o$C). In this case the growth temperature was $\sim$500$^o$C which resulted in lower stress.

Raman spectra for LTO, SiO$_2$ coated LTO (SLTO) and diamond film grown on SLTO were also taken, similar to the data on LNO, and the results are presented in figure \ref{fig5}. Like before, the bottom spectra (shown in blue solid line) is for bare substrate. The crystal shows the characteristic features for a x-cut LTO crystal\cite{pen76, rep99, shur14}. The observed peaks from the literature\cite{pen76, rep99} for LTO are 140, 180, 201, 206, 251, 253, 316, 356. 383, 462, 597, 660 and 750 cm$^{-1}$. In the data presented in figure \ref{fig5}, peaks are seen at $\sim$ 142, 191, 207, 252, 318, 358, 384, 463, 597, 659 and 751. The peak at 191 cm$^{-1}$ is close to the calculated frequency of 199.9 cm$^{-1}$\cite{rep99}. This frequency has been reported to appear at 180 cm$^{-1}$ in many studies.  The peak at 597 cm$^{-1}$ generally include the calculated peaks at 595, 599.4 and 617 cm$^{-1}$\cite{rep99}. The curve in green dashed in figure \ref{fig5} shows the Raman spectra for SLTO. The relative intensities of various peaks from the substrate are markedly different, as has been seen in the case of LNO. Finally the curve in red dotted line shows the spectrum for diamond grown on SLTO. The characteristic peaks for LTO are all present and as seen for diamond on SLNO, the signal from the diamond layer is weak and can only be seen in the zoomed in version shown in the inset. Similar to LNO and SLNO, no such peaks were seen in the LTO or SLTO spectra and hence the expanded view have not been shown for the same. The diamond peak position is $\sim$1330.84 cm$^{-1}$ as determined by a Gaussian fit. The stress in the diamond film on SLNO is around $\sim$0.46 GPa\cite{bop85, knig89, anas99}.

\section{Conclusion}
In conclusion, we have measured the $\zeta$-potential of LNO and LTO single crystal and studied electrostatic driven self assembly of nanodiamonds on these crystal. The $\zeta$-potential of both the substrates are negative. The presence of oxygen groups, as determined by XPS, on the surface are most likely reason for the negative $\zeta$-potential. The self assembled nanodiamonds were used to grow thin diamond layers on these substrates. It was found that the substrates could not survive the harsh diamond growth condition. As a result a protective SiO$_2$ layer was deposited for diamond growth experiments. It has been shown here that it is possible to successfully grow thin diamond layers on buffered LNO and LTO. However, the growth rate in both the cases is really slow (around 40nm/hr). This growth rate is not ideal for growing tens of microns of diamond needed for effective thermal management in acoustic wave devices. Furthermore, it is important to have large grain size to benefit from good thermal property of diamond and also to reduce losses in acoustic wave filters. Finally, even if diamond is grown at such slow growth rates to get thicker diamond layers, the layers peel-off from the substrate in-spite of low stress seen in thinner layers. It is quite possible that as the diamond layer grows thicker the stress between diamond and the substrate also increases. The thin diamond layers were analysed with AFM and Raman spectroscopy and were found to be of good crystallinity with minimal stress. The alternative to this would be to directly bond piezo material to poly-diamond substrates which has its own complexities due the requirement of low surface roughnesses ($<$1nm over large area).

\section*{Acknowledgment}
The authors would like to acknowledge financial support from Huawei.
\bibliography{cite}

\begin{thebibliography}{10}
\expandafter\ifx\csname url\endcsname\relax
  \def\url#1{\texttt{#1}}\fi
\expandafter\ifx\csname urlprefix\endcsname\relax\def\urlprefix{URL }\fi
\expandafter\ifx\csname href\endcsname\relax
  \def\href#1#2{#2} \def\path#1{#1}\fi

\bibitem{dels19}
P.~Delsing, A.~N. Cleland, M.~J. Schuetz, J.~Kn{\"{o}}rzer, G.~Giedke, J.~I.
  Cirac, K.~Srinivasan, M.~Wu, K.~C. Balram, C.~Ba{\"{u}}erle, T.~Meunier,
  C.~J. Ford, P.~V. Santos, E.~Cerda-M{\'{e}}ndez, H.~Wang, H.~J. Krenner,
  E.~D. Nysten, M.~Wei{\ss}, G.~R. Nash, L.~Thevenard, C.~Gourdon,
  P.~Rovillain, M.~Marangolo, J.~Y. Duquesne, G.~Fischerauer, W.~Ruile,
  A.~Reiner, B.~Paschke, D.~Denysenko, D.~Volkmer, A.~Wixforth, H.~Bruus,
  M.~Wiklund, J.~Reboud, J.~M. Cooper, Y.~Q. Fu, M.~S. Brugger, F.~Rehfeldt,
  C.~Westerhausen, {The 2019 surface acoustic waves roadmap}, Journal of
  Physics D: Applied Physics 52~(35) (2019).
\newblock \href {https://doi.org/10.1088/1361-6463/ab1b04}
  {\path{doi:10.1088/1361-6463/ab1b04}}.

\bibitem{ieee}
\href{https://futurenetworks.ieee.org/images/files/pdf/ieee-5g-roadmap-white-paper.pdf}{{IEEE
  Future Networks Technology Roadmap Working Group 2017 IEEE 5G and beyond
  technology roadmap white paper}}, Tech. rep. (2017).
\newline\urlprefix\url{https://futurenetworks.ieee.org/images/files/pdf/ieee-5g-roadmap-white-paper.pdf}

\bibitem{bal22}
O.~L. Balysheva, Saw filters substrates for 5g filters, in: 2022 Wave
  Electronics and its Application in Information and Telecommunication Systems
  (WECONF), 2022, pp. 1--7.
\newblock \href {https://doi.org/10.1109/WECONF55058.2022.9803545}
  {\path{doi:10.1109/WECONF55058.2022.9803545}}.

\bibitem{shen22}
J.~Shen, S.~Fu, R.~Su, H.~Xu, Z.~Lu, Q.~Zhang, F.~Zeng, C.~Song, W.~Wang,
  F.~Pan, {SAW Filters With Excellent Temperature Stability and High Power
  Handling Using LiTaO 3 /SiC Bonded Wafers}, Journal of Microelectromechanical
  Systems 31~(2) (2022) 186--193.
\newblock \href {https://doi.org/10.1109/JMEMS.2021.3137928}
  {\path{doi:10.1109/JMEMS.2021.3137928}}.

\bibitem{iva02}
E.~V. Ivakin, A.~V. Sukhodolov, V.~G. Ralchenko, A.~V. Vlasov, A.~V. Khomich,
  {Measurement of thermal conductivity of polycrystalline CVD diamond by
  laser-induced transient grating technique}, Quantum Electronics 32~(4) (2002)
  367--372.
\newblock \href {https://doi.org/10.1070/QE2002v032n04ABEH002200}
  {\path{doi:10.1070/QE2002v032n04ABEH002200}}.

\bibitem{hee13}
J.~Hees, N.~Heidrich, W.~Pletschen, R.~E. Sah, M.~Wolfer, O.~A. Williams,
  V.~Lebedev, C.~E. Nebel, O.~Ambacher, {Piezoelectric actuated
  micro-resonators based on the growth of diamond on aluminum nitride thin
  films}, Nanotechnology 24~(2) (2013) 025601.
\newblock \href {https://doi.org/10.1088/0957-4484/24/2/025601}
  {\path{doi:10.1088/0957-4484/24/2/025601}}.

\bibitem{man19}
S.~Mandal, C.~Yuan, F.~Massabuau, J.~W. Pomeroy, J.~Cuenca, H.~Bland,
  E.~Thomas, D.~Wallis, T.~Batten, D.~Morgan, R.~Oliver, M.~Kuball, O.~A.
  Williams, {Thick, Adherent Diamond Films on AlN with Low Thermal Barrier
  Resistance}, ACS Applied Materials \& Interfaces 11~(43) (2019) 40826--40834.
\newblock \href {https://doi.org/10.1021/acsami.9b13869}
  {\path{doi:10.1021/acsami.9b13869}}.

\bibitem{tak99}
H.~Takagi, R.~Maeda, N.~Hosoda, T.~Suga, {Room-temperature bonding of lithium
  niobate and silicon wafers by argon-beam surface activation}, Applied Physics
  Letters 74~(16) (1999) 2387--2389.
\newblock \href {https://doi.org/10.1063/1.123860}
  {\path{doi:10.1063/1.123860}}.

\bibitem{tor08}
I.~Torchinsky, G.~Rosenman, {Interface Engineering and Direct Bonding of
  Lithium Tantalate Crystals}, Journal of Electronic Materials 37~(10) (2008)
  1547--1551.
\newblock \href {https://doi.org/10.1007/s11664-008-0508-2}
  {\path{doi:10.1007/s11664-008-0508-2}}.

\bibitem{mu18}
F.~Mu, R.~He, T.~Suga, {Room temperature GaN-diamond bonding for high-power
  GaN-on-diamond devices}, Scripta Materialia 150 (2018) 148--151.
\newblock \href {https://doi.org/10.1016/j.scriptamat.2018.03.016}
  {\path{doi:10.1016/j.scriptamat.2018.03.016}}.

\bibitem{lian19}
J.~Liang, S.~Yamajo, M.~Kuball, N.~Shigekawa, {Room-temperature direct bonding
  of diamond and Al}, Scripta Materialia 159 (2019) 58--61.
\newblock \href {https://doi.org/10.1016/j.scriptamat.2018.09.016}
  {\path{doi:10.1016/j.scriptamat.2018.09.016}}.

\bibitem{mats20}
T.~Matsumae, Y.~Kurashima, H.~Umezawa, K.~Tanaka, T.~Ito, H.~Watanabe,
  H.~Takagi, {Low-temperature direct bonding of $\beta$-Ga 2 O 3 and diamond
  substrates under atmospheric conditions}, Applied Physics Letters 116~(14)
  (2020) 141602.
\newblock \href {https://doi.org/10.1063/5.0002068}
  {\path{doi:10.1063/5.0002068}}.

\bibitem{rod12}
J.~G. Rodriguez-Madrid, G.~F. Iriarte, J.~Pedros, O.~A. Williams, D.~Brink,
  F.~Calle, {Super-High-Frequency SAW Resonators on AlN/Diamond}, IEEE Electron
  Device Letters 33~(4) (2012) 495--497.
\newblock \href {https://doi.org/10.1109/LED.2012.2183851}
  {\path{doi:10.1109/LED.2012.2183851}}.

\bibitem{mig22}
M.~{Sinusia Lozano}, L.~Fern{\'{a}}ndez-Garc{\'{i}}a, D.~L{\'{o}}pez-Romero,
  O.~A. Williams, G.~F. Iriarte, {SAW Resonators and Filters Based on
  Sc0.43Al0.57N on Single Crystal and Polycrystalline Diamond}, Micromachines
  13~(7) (2022) 1--9.
\newblock \href {https://doi.org/10.3390/mi13071061}
  {\path{doi:10.3390/mi13071061}}.

\bibitem{zha21}
H.~Zhang, Z.~Guo, {Thickness Dependence and Anisotropy of Capped Diamond
  Thermal Conductivity on Cooling of Pulse-Operated GaN HEMTs}, IEEE
  Transactions on Components, Packaging and Manufacturing Technology 11~(2)
  (2021) 233--240.
\newblock \href {https://doi.org/10.1109/TCPMT.2021.3050976}
  {\path{doi:10.1109/TCPMT.2021.3050976}}.

\bibitem{hark42}
W.~D. Harkins, {Energy Relations of the Surface of Solids I. Surface Energy of
  the Diamond}, The Journal of Chemical Physics 10~(5) (1942) 268--272.
\newblock \href {https://doi.org/10.1063/1.1723719}
  {\path{doi:10.1063/1.1723719}}.

\bibitem{jacc63}
R.~J. Jaccodine, {Surface Energy of Germanium and Silicon}, Journal of The
  Electrochemical Society 110~(6) (1963) 524.
\newblock \href {https://doi.org/10.1149/1.2425806}
  {\path{doi:10.1149/1.2425806}}.

\bibitem{will11rev}
O.~Williams, {Nanocrystalline diamond}, Diamond and Related Materials 20~(5-6)
  (2011) 621--640.
\newblock \href {https://doi.org/10.1016/j.diamond.2011.02.015}
  {\path{doi:10.1016/j.diamond.2011.02.015}}.

\bibitem{mandrev}
S.~Mandal, {Nucleation of diamond films on heterogeneous substrates: a review},
  RSC Advances 11 (2021) 10159--10182.
\newblock \href {https://doi.org/10.1039/D1RA00397F}
  {\path{doi:10.1039/D1RA00397F}}.

\bibitem{hir20}
Y.~Hirsh, S.~Gorfman, D.~Sherman, {Cleavage and surface energies of LiNbO3},
  Acta Materialia 193 (2020) 338--349.
\newblock \href {https://doi.org/10.1016/j.actamat.2020.03.046}
  {\path{doi:10.1016/j.actamat.2020.03.046}}.

\bibitem{obr30}
J.~W. Obreimoff, {The splitting strength of mica}, Proceedings of the Royal
  Society of London A 127~(805) (1930) 290--297.
\newblock \href {https://doi.org/10.1098/rspa.1930.0058}
  {\path{doi:10.1098/rspa.1930.0058}}.

\bibitem{bak19}
B.~Baker, N.~Herbots, S.~D. Whaley, M.~Sahal, J.~Kintz, A.~Yano, S.~Narayan,
  A.~L. Brimhall, W.-L. Lee, Y.~Akabane, R.~J. Culbertson, {Surface energy
  engineering for LiTaO 3 and $\alpha$-quartz SiO 2 for low temperature ($<$220
  °C) wafer bonding}, Journal of Vacuum Science \& Technology A 37~(4) (2019)
  041101.
\newblock \href {https://doi.org/10.1116/1.5095157}
  {\path{doi:10.1116/1.5095157}}.

\bibitem{man16}
S.~Mandal, E.~L.~H. Thomas, T.~A. Jenny, O.~A. Williams, {Chemical Nucleation
  of Diamond Films}, ACS Applied Materials \& Interfaces 8~(39) (2016)
  26220--26225.
\newblock \href {https://doi.org/10.1021/acsami.6b08286}
  {\path{doi:10.1021/acsami.6b08286}}.

\bibitem{will07s}
O.~A. Williams, O.~Douh\'{e}ret, M.~Daenen, K.~Haenen, E.~\={O}sawa,
  M.~Takahashi, Enhanced diamond nucleation on monodispersed nanocrystalline
  diamond, Chemical Physics Letters 445~(4–6) (2007) 255 -- 258.
\newblock \href {https://doi.org/10.1016/j.cplett.2007.07.091}
  {\path{doi:10.1016/j.cplett.2007.07.091}}.

\bibitem{hee11}
J.~Hees, A.~Kriele, O.~A. Williams, {Electrostatic self-assembly of diamond
  nanoparticles}, Chemical Physics Letters 509~(1-3) (2011) 12--15.
\newblock \href {https://doi.org/10.1016/j.cplett.2011.04.083}
  {\path{doi:10.1016/j.cplett.2011.04.083}}.

\bibitem{lev66}
H.~J. Levinstein, A.~A. Ballman, C.~D. Capio, {Domain Structure and Curie
  Temperature of Single‐Crystal Lithium Tantalate}, Journal of Applied
  Physics 37~(12) (1966) 4585--4586.
\newblock \href {https://doi.org/10.1063/1.1708088}
  {\path{doi:10.1063/1.1708088}}.

\bibitem{tur98}
H.~Turcicov{\'{a}}, V.~Perina, J.~Vac{\'{i}}k, J.~Cerven{\'{a}}, J.~Zemek,
  V.~Zelezn{\'{y}}, H.~Arend, {Plasma processing of in a hydrogen/oxygen
  radio-frequency discharge}, Journal of Physics D: Applied Physics 31~(9)
  (1998) 1052--1059.
\newblock \href {https://doi.org/10.1088/0022-3727/31/9/004}
  {\path{doi:10.1088/0022-3727/31/9/004}}.

\bibitem{kim69}
Y.~S. Kim, R.~T. Smith, {Thermal Expansion of Lithium Tantalate and Lithium
  Niobate Single Crystals}, Journal of Applied Physics 40~(11) (1969)
  4637--4641.
\newblock \href {https://doi.org/10.1063/1.1657244}
  {\path{doi:10.1063/1.1657244}}.

\bibitem{moel97}
C.~Moelle, S.~Klose, F.~Sz{\"{u}}cs, H.~Fecht, C.~Johnston, P.~Chalker,
  M.~Werner, {Measurement and calculation of the thermal expansion coefficient
  of diamond}, Diamond and Related Materials 6~(5-7) (1997) 839--842.
\newblock \href {https://doi.org/10.1016/S0925-9635(96)00674-7}
  {\path{doi:10.1016/S0925-9635(96)00674-7}}.

\bibitem{man21}
S.~Mandal, K.~Arts, H.~C. Knoops, J.~A. Cuenca, G.~M. Klemencic, O.~A.
  Williams, Surface zeta potential and diamond growth on gallium oxide single
  crystal, Carbon 181 (2021) 79--86.
\newblock \href {https://doi.org/10.1016/j.carbon.2021.04.100}
  {\path{doi:10.1016/j.carbon.2021.04.100}}.

\bibitem{jag04}
K.~Jagannadham, M.~J. Lance, T.~R. Watkins, {Growth of diamond film on single
  crystal lithium niobate for surface acoustic wave devices}, Journal of Vacuum
  Science \& Technology A: Vacuum, Surfaces, and Films 22~(4) (2004)
  1105--1109.
\newblock \href {https://doi.org/10.1116/1.1740770}
  {\path{doi:10.1116/1.1740770}}.

\bibitem{vor11}
A.~Vorobiev, J.~Berge, S.~Gevorgian, M.~L{\"{o}}ffler, E.~Olsson, {Effect of
  interface roughness on acoustic loss in tunable thin film bulk acoustic wave
  resonators}, Journal of Applied Physics 110~(2) (2011) 024116.
\newblock \href {https://doi.org/10.1063/1.3610513}
  {\path{doi:10.1063/1.3610513}}.

\bibitem{wag80}
R.~{Van Wagenen}, J.~Andrade, {Flat plate streaming potential investigations:
  Hydrodynamics and electrokinetic equivalency}, Journal of Colloid and
  Interface Science 76~(2) (1980) 305--314.
\newblock \href {https://doi.org/10.1016/0021-9797(80)90374-4}
  {\path{doi:10.1016/0021-9797(80)90374-4}}.

\bibitem{wern98}
C.~Werner, H.~K{\"{o}}rber, R.~Zimmermann, S.~Dukhin, H.-J. Jacobasch,
  {Extended Electrokinetic Characterization of Flat Solid Surfaces}, Journal of
  Colloid and Interface Science 208~(1) (1998) 329--346.
\newblock \href {https://doi.org/10.1006/jcis.1998.5787}
  {\path{doi:10.1006/jcis.1998.5787}}.

\bibitem{will10h}
O.~A. Williams, J.~Hees, C.~Dieker, W.~J\"{a}ger, L.~Kirste, C.~E. Nebel,
  Size-dependent reactivity of diamond nanoparticles, ACS Nano 4~(8) (2010)
  4824--4830.
\newblock \href {https://doi.org/10.1021/nn100748k}
  {\path{doi:10.1021/nn100748k}}.

\bibitem{blan21}
H.~A. Bland, I.~A. Centeleghe, S.~Mandal, E.~L.~H. Thomas, J.-y. Maillard,
  O.~A. Williams, {Electropositive Nanodiamond-Coated Quartz Microfiber
  Membranes for Virus and Dye Filtration}, ACS Applied Nano Materials 4~(3)
  (2021) 3252--3261.
\newblock \href {https://doi.org/10.1021/acsanm.1c00439}
  {\path{doi:10.1021/acsanm.1c00439}}.

\bibitem{heil07}
S.~B.~S. Heil, J.~L. van Hemmen, C.~J. Hodson, N.~Singh, J.~H. Klootwijk,
  F.~Roozeboom, M.~C.~M. van~de Sanden, W.~M.~M. Kessels, {Deposition of TiN
  and HfO[sub 2] in a commercial 200 mm remote plasma atomic layer deposition
  reactor}, Journal of Vacuum Science {\&} Technology A: Vacuum, Surfaces, and
  Films 25~(5) (2007) 1357.
\newblock \href {https://doi.org/10.1116/1.2753846}
  {\path{doi:10.1116/1.2753846}}.

\bibitem{lang09}
E.~Langereis, S.~B.~S. Heil, H.~C.~M. Knoops, W.~Keuning, M.~C.~M. van~de
  Sanden, W.~M.~M. Kessels, {In situ spectroscopic ellipsometry as a versatile
  tool for studying atomic layer deposition}, Journal of Physics D: Applied
  Physics 42~(7) (2009) 073001.
\newblock \href {https://doi.org/10.1088/0022-3727/42/7/073001}
  {\path{doi:10.1088/0022-3727/42/7/073001}}.

\bibitem{man17}
S.~Mandal, E.~L.~H. Thomas, C.~Middleton, L.~Gines, J.~T. Griffiths, M.~J.
  Kappers, R.~A. Oliver, D.~J. Wallis, L.~E. Goff, S.~A. Lynch, M.~Kuball,
  O.~A. Williams, {Surface Zeta Potential and Diamond Seeding on Gallium
  Nitride Films}, ACS Omega 2~(10) (2017) 7275--7280.
\newblock \href {https://doi.org/10.1021/acsomega.7b01069}
  {\path{doi:10.1021/acsomega.7b01069}}.

\bibitem{cuen22}
J.~A. Cuenca, S.~Mandal, E.~L. Thomas, O.~A. Williams, Microwave plasma
  modelling in clamshell chemical vapour deposition diamond reactors, Diamond
  and Related Materials 124 (2022) 108917.
\newblock \href {https://doi.org/10.1016/j.diamond.2022.108917}
  {\path{doi:10.1016/j.diamond.2022.108917}}.

\bibitem{cuen21}
J.~A. Cuenca, M.~D. Smith, D.~E. Field, F.~{C-P. Massabuau}, S.~Mandal,
  J.~Pomeroy, D.~J. Wallis, R.~A. Oliver, I.~Thayne, M.~Kuball, O.~A. Williams,
  {Thermal stress modelling of diamond on GaN/III-Nitride membranes}, Carbon
  174 (2021) 647--661.
\newblock \href {https://doi.org/10.1016/j.carbon.2020.11.067}
  {\path{doi:10.1016/j.carbon.2020.11.067}}.

\bibitem{fair21}
N.~Fairley, V.~Fernandez, M.~Richard‐Plouet, C.~Guillot-Deudon, J.~Walton,
  E.~Smith, D.~Flahaut, M.~Greiner, M.~Biesinger, S.~Tougaard, D.~Morgan,
  J.~Baltrusaitis, {Systematic and collaborative approach to problem solving
  using X-ray photoelectron spectroscopy}, Applied Surface Science Advances
  5~(June) (2021) 100112.
\newblock \href {https://doi.org/10.1016/j.apsadv.2021.100112}
  {\path{doi:10.1016/j.apsadv.2021.100112}}.

\bibitem{gira09}
H.~A. Girard, S.~Perruchas, C.~Gesset, M.~Chaigneau, L.~Vieille, J.~C. Arnault,
  P.~Bergonzo, J.~P. Boilot, T.~Gacoin, {Electrostatic grafting of diamond
  nanoparticles: A versatile route to nanocrystalline diamond thin films}, ACS
  Applied Materials and Interfaces 1~(12) (2009) 2738--2746.
\newblock \href {https://doi.org/10.1021/am900458g}
  {\path{doi:10.1021/am900458g}}.

\bibitem{scor09}
E.~Scorsone, S.~Saada, J.~C. Arnault, P.~Bergonzo, {Enhanced control of diamond
  nanoparticle seeding using a polymer matrix}, Journal of Applied Physics
  106~(1) (2009) 2--8.
\newblock \href {https://doi.org/10.1063/1.3153118}
  {\path{doi:10.1063/1.3153118}}.

\bibitem{hunt81}
R.~J. HUNTER, Chapter 8 - influence of more complex adsorbates on zeta
  potential, in: R.~J. HUNTER (Ed.), Zeta Potential in Colloid Science,
  Academic Press, 1981, pp. 305--344.
\newblock \href
  {https://doi.org/https://doi.org/10.1016/B978-0-12-361961-7.50012-2}
  {\path{doi:https://doi.org/10.1016/B978-0-12-361961-7.50012-2}}.

\bibitem{wilso01}
J.~Wilson, J.~Walton, G.~Beamson, {Analysis of Chemical Vapour Deposited
  Diamond Films by X-ray Photoelectron Spectroscopy}, Journal of Electron
  Spectroscopy and Related Phenomena 121~(1-3) (2001) 183--201.
\newblock \href {https://doi.org/10.1016/S0368-2048(01)00334-6}
  {\path{doi:10.1016/S0368-2048(01)00334-6}}.

\bibitem{ferr05}
S.~Ferro, M.~{Dal Colle}, A.~{De Battisti}, {Chemical Surface Characterization
  of Electrochemically and Thermally Oxidized Boron-doped Diamond Film
  Electrodes}, Carbon 43~(6) (2005) 1191--1203.
\newblock \href {https://doi.org/10.1016/j.carbon.2004.12.012}
  {\path{doi:10.1016/j.carbon.2004.12.012}}.

\bibitem{skry16}
E.~Skryleva, I.~Kubasov, P.~Kiryukhantsev-Korneev, B.~Senatulin, R.~Zhukov,
  K.~Zakutailov, M.~Malinkovich, Y.~Parkhomenko, {XPS study of Li/Nb ratio in
  LiNbO3 crystals. Effect of polarity and mechanical processing on LiNbO3
  surface chemical composition}, Applied Surface Science 389 (2016) 387--394.
\newblock \href {https://doi.org/10.1016/j.apsusc.2016.07.108}
  {\path{doi:10.1016/j.apsusc.2016.07.108}}.

\bibitem{gitm95}
F.~Gitmans, Z.~Sitar, P.~G{\"{u}}nter, {Growth of tantalum oxide and lithium
  tantalate thin films by molecular beam epitaxy}, Vacuum 46~(8-10) (1995)
  939--942.
\newblock \href {https://doi.org/10.1016/0042-207X(95)00077-1}
  {\path{doi:10.1016/0042-207X(95)00077-1}}.

\bibitem{ono02}
S.~Ono, S.-i. Hirano, {Processing of highly oriented lithium tantalate films by
  chemical solution deposition}, Journal of Materials Research 17~(10) (2002)
  2532--2539.
\newblock \href {https://doi.org/10.1557/JMR.2002.0368}
  {\path{doi:10.1557/JMR.2002.0368}}.

\bibitem{yun07}
Y.~Yun, M.~Li, D.~Liao, L.~Kampschulte, E.~Altman, {Geometric and electronic
  structure of positively and negatively poled LiNbO3 (0 0 0 1) surfaces},
  Surface Science 601~(19) (2007) 4636--4647.
\newblock \href {https://doi.org/10.1016/j.susc.2007.08.001}
  {\path{doi:10.1016/j.susc.2007.08.001}}.

\bibitem{nots99}
H.~Notsu, I.~Yagi, T.~Tatsuma, D.~A. Tryk, A.~Fujishima, {Introduction of
  Oxygen-Containing Functional Groups onto Diamond Electrode Surfaces by Oxygen
  Plasma and Anodic Polarization}, Electrochemical and Solid-State Letters
  2~(10) (1999) 522.
\newblock \href {https://doi.org/10.1149/1.1390890}
  {\path{doi:10.1149/1.1390890}}.

\bibitem{nots00}
H.~Notsu, I.~Yagi, T.~Tatsuma, D.~A. Tryk, A.~Fujishima, {Surface carbonyl
  groups on oxidized diamond electrodes}, Journal of Electroanalytical
  Chemistry 492~(1) (2000) 31--37.
\newblock \href {https://doi.org/10.1016/S0022-0728(00)00254-0}
  {\path{doi:10.1016/S0022-0728(00)00254-0}}.

\bibitem{zaj01}
L.~Zaj{\'{i}}{\v{c}}kov{\'{a}}, K.~Veltrusk{\'{a}}, N.~Tsud, D.~Franta, {XPS
  and ellipsometric study of DLC/silicon interface}, Vacuum 61~(2-4) (2001)
  269--273.
\newblock \href {https://doi.org/10.1016/S0042-207X(01)00128-2}
  {\path{doi:10.1016/S0042-207X(01)00128-2}}.

\bibitem{gard95}
S.~D. Gardner, C.~S. Singamsetty, G.~L. Booth, G.-R. He, C.~U. Pittman,
  {Surface characterization of carbon fibers using angle-resolved XPS and ISS},
  Carbon 33~(5) (1995) 587--595.
\newblock \href {https://doi.org/10.1016/0008-6223(94)00144-O}
  {\path{doi:10.1016/0008-6223(94)00144-O}}.

\bibitem{zhen20}
Q.~Zheng, Y.~Zhang, {Microstructure and dielectric properties of LiTaO3
  ceramics with MnO2 addition fabricated by hot-pressing sintering}, Applied
  Physics A 126~(2) (2020) 104.
\newblock \href {https://doi.org/10.1007/s00339-020-3281-6}
  {\path{doi:10.1007/s00339-020-3281-6}}.

\bibitem{vol26}
M.~Volmer, A.~Weber, {Keimbildung in {\"{u}}bers{\"{a}}ttigten Gebilden},
  Zeitschrift f{\"{u}}r Physikalische Chemie 119U~(1) (1926) 277--301.
\newblock \href {https://doi.org/10.1515/zpch-1926-11927}
  {\path{doi:10.1515/zpch-1926-11927}}.

\bibitem{jia94}
X.~Jiang, K.~Schiffmann, C.-P.~P. Klages, {Nucleation and initial growth phase
  of diamond thin films on (100) silicon}, Physical Review B 50~(12) (1994)
  8402--8410.
\newblock \href {https://doi.org/10.1103/PhysRevB.50.8402}
  {\path{doi:10.1103/PhysRevB.50.8402}}.

\bibitem{van67}
A.~{Van der Drift}, {Evolutionary selection, a principle governing growth
  orientation in vapour-deposited layers}, Philips Res. Rep 22~(3) (1967) 267.

\bibitem{smer05}
P.~Smereka, X.~Li, G.~Russo, D.~Srolovitz, {Simulation of faceted film growth
  in three dimensions: microstructure, morphology and texture}, Acta Materialia
  53~(4) (2005) 1191--1204.
\newblock \href {https://doi.org/10.1016/j.actamat.2004.11.013}
  {\path{doi:10.1016/j.actamat.2004.11.013}}.

\bibitem{sch66}
R.~F. Schaufele, M.~J. Weber, {Raman Scattering by Lithium Niobate}, Physical
  Review 152~(2) (1966) 705--708.
\newblock \href {https://doi.org/10.1103/PhysRev.152.705}
  {\path{doi:10.1103/PhysRev.152.705}}.

\bibitem{rep99}
Y.~Repelin, E.~Husson, F.~Bennani, C.~Proust, {Raman spectroscopy of lithium
  niobate and lithium tantalate. Force field calculations}, Journal of Physics
  and Chemistry of Solids 60~(6) (1999) 819--825.
\newblock \href {https://doi.org/10.1016/S0022-3697(98)00333-3}
  {\path{doi:10.1016/S0022-3697(98)00333-3}}.

\bibitem{ram30}
C.~Ramaswamy, {Raman Effect in Diamond}, Nature 125 (1930) 704--704.
\newblock \href {https://doi.org/10.1038/125704b0}
  {\path{doi:10.1038/125704b0}}.

\bibitem{bhag30}
S.~Bhagavantam, \href{http://hdl.handle.net/10821/555}{{Relation of Raman
  Effect to Crystal Structure}}, Indian Journal of Physics 5 (1930) 169.
\newline\urlprefix\url{http://hdl.handle.net/10821/555}

\bibitem{bop85}
H.~Boppart, J.~van Straaten, I.~F. Silvera, {Raman spectra of diamond at high
  pressures}, Physical Review B 32~(2) (1985) 1423--1425.
\newblock \href {https://doi.org/10.1103/PhysRevB.32.1423}
  {\path{doi:10.1103/PhysRevB.32.1423}}.

\bibitem{knig89}
D.~S. Knight, W.~B. White, {Characterization of diamond films by Raman
  spectroscopy}, Journal of Materials Research 4~(2) (1989) 385--393.
\newblock \href {https://doi.org/10.1557/JMR.1989.0385}
  {\path{doi:10.1557/JMR.1989.0385}}.

\bibitem{anas99}
E.~Anastassakis, {Strain characterization of polycrystalline diamond and
  silicon systems}, Journal of Applied Physics 86~(1) (1999) 249--258.
\newblock \href {https://doi.org/10.1063/1.370723}
  {\path{doi:10.1063/1.370723}}.

\bibitem{pen76}
A.~F. Penna, A.~Chaves, P.~D.~R. Andrade, S.~P.~S. Porto, {Light scattering by
  lithium tantalate at room temperature}, Physical Review B 13~(11) (1976)
  4907--4917.
\newblock \href {https://doi.org/10.1103/PhysRevB.13.4907}
  {\path{doi:10.1103/PhysRevB.13.4907}}.

\bibitem{shur14}
V.~{Ya. Shur}, P.~S. Zelenovskiy, {Micro- and nanodomain imaging in uniaxial
  ferroelectrics: Joint application of optical, confocal Raman, and
  piezoelectric force microscopy}, Journal of Applied Physics 116~(6) (2014)
  066802.
\newblock \href {https://doi.org/10.1063/1.4891397}
  {\path{doi:10.1063/1.4891397}}.

\end{thebibliography}
\end{document}